\documentclass[10pt]{wlscirep}
\usepackage{multirow}
\usepackage{makecell}
\usepackage{graphicx}
\usepackage[label font=bf,labelformat=simple]{subfig}% <-- changed
\usepackage{floatrow}
\usepackage{color}
\usepackage{comment}
\usepackage{lineno}
\title{Robust haplotype-resolved assembly of diploid individuals without parental data}

\author[1,2]{Haoyu Cheng}
\author[3,4]{Erich D. Jarvis}
\author[3]{Olivier Fedrigo}
\author[5,6,7]{Klaus-Peter Koepfli}
\author[8]{Lara Urban}
\author[8]{Neil J. Gemmell}
\author[1,2,*]{Heng Li}
\affil[1]{Department of Data Sciences, Dana-Farber Cancer Institute, Boston, MA, USA}
\affil[2]{Department of Biomedical Informatics, Harvard Medical School, Boston, MA, USA}
\affil[3]{The Vertebrate Genome Lab, The Rockefeller University, New York, NY 10065}
\affil[4]{Howard Hughes Medical Institute, Chevy Chase, MD, 20815}
\affil[5]{Smithsonian-Mason School of Conservation, George Mason University, Front Royal, VA 22630, USA}
\affil[6]{Smithsonian Conservation Biology Institute, Center for Species Survival, National Zoological Park, Washington, D.C., 20008, USA}
\affil[7]{ITMO University, Computer Technologies Laboratory, St. Petersburg 197101, Russia}
\affil[8]{Department of Anatomy, University of Otago, Dunedin 9016, New Zealand}
\affil[*]{To whom correspondence should be addressed: hli@ds.dfci.harvard.edu}

\begin{abstract}
Routine single-sample haplotype-resolved assembly remains an unresolved problem.
Here we describe a new algorithm that combines PacBio HiFi reads and Hi-C chromatin interaction data
to produce a haplotype-resolved assembly without the sequencing of parents.
Applied to human and other vertebrate samples,
our algorithm consistently outperforms existing single-sample assembly pipelines
and generates assemblies of comparable quality to the best pedigree-based assemblies.
\end{abstract}

\begin{document}

\maketitle

The advances in long-read sequencing technologies have laid the foundations for high-quality \emph{de novo} genome assembly~\cite{logsdon2020long, rhie2021towards}.
For diploid or polyploid genomes, most long-read assemblers collapse different homologous haplotypes into a consensus assembly,
completely losing the phasing information.
A few assemblers use heterozygous differences between haplotypes to retain local
phasing~\cite{chin2016phased,nurk2020hicanu,cheng2021haplotype}.
However, owing to the limited read length,
these assemblers can only generate short phased blocks for samples of low heterozygosity.
To obtain long contigs,
they stitch part of these short blocks to produce a primary assembly with homologous haplotypes randomly switching in each contig,
and dump the remaining blocks into an alternate assembly.
The alternate assembly is fragmented and often contains excessive sequence duplications. It is ignored in most downstream analysis.
To this end, a primary/alternate assembly still represents a single haplotype, not a diploid genome.
For a diploid genome, we prefer to derive a \emph{haplotype-resolved assembly}
which consists of two sets of contigs with each set representing a complete homologous haplotype.
Each contig in a haplotype-resolved assembly supposedly comes from one haplotype, also known as a \emph{haplotig}.
Trio binning~\cite{koren2018novo} is the first algorithm to produce a haplotype-resolved assembly
but the requirement of parental sequencing often limits its application in practice.
This limitation motivates the recent development of single-sample haplotype-resolved assembly using additional Hi-C or
Strand-Seq data~\cite{garg2021chromosome,porubsky2021fully,kronenberg2021extended}.
These methods all start with a collapsed assembly and ignore the rich information in phased assembly graphs.
They are not as good as graph trio binning employed by hifiasm~\cite{cheng2021haplotype}.
In addition, the published single-sample methods all involve multiple steps in a long pipeline,
which complicates their installation and deployment.

To overcome the limitations of earlier methods,
we extended our hifiasm assembler to single-sample haplotype-resolved assembly using HiFi and Hi-C data both at around 30-fold coverage.
Our Hi-C based algorithm is built on top of phased hifiasm assembly graphs~\cite{cheng2021haplotype}
but differs from the original hifiasm in sequence parition.
In a hifiasm graph, each node is a unitig assembled from HiFi reads with correct phasing and each edge represents an overlap between two unitigs.
While the old algorithm labels reads in unitigs with parental k-mers,
the new algorithm partitions relatively short unitigs in the graph with Hi-C short reads (Fig.~1a).
In more detail, we index 31-mers in unitigs and map Hi-C short reads to them without detailed base alignment.
If a Hi-C fragment harbors two distant heterozygotes on two unitigs, we add a ``link'' between the unitigs.
This provides long-range phasing information.
We then bipartition the unitigs such that unitigs in each partition have little redundancy and share many Hi-C links.
We reduce such unitig bipartition to a graph max-cut problem~\cite{Edge:2017aa}
and find a near optimal solution with a stochastic algorithm~\cite{Tourdot:2021aa} in the principle of simulated annealing (Online Methods).
We also consider the topology of the assembly graph to reduce the chance of local optima.
In the end, we reuse the same graph binning strategy in the original hifiasm to produce the final haplotype-resolved assembly.
Unlike existing methods, our algorithm directly operates on a HiFi assembly graph and tightly integrates Hi-C read mapping,
phasing and assembly into one single executable program with no dependency to external tools.
It is easier to use and runs faster.

\begin{figure}[!tp]
\includegraphics[width=\textwidth]{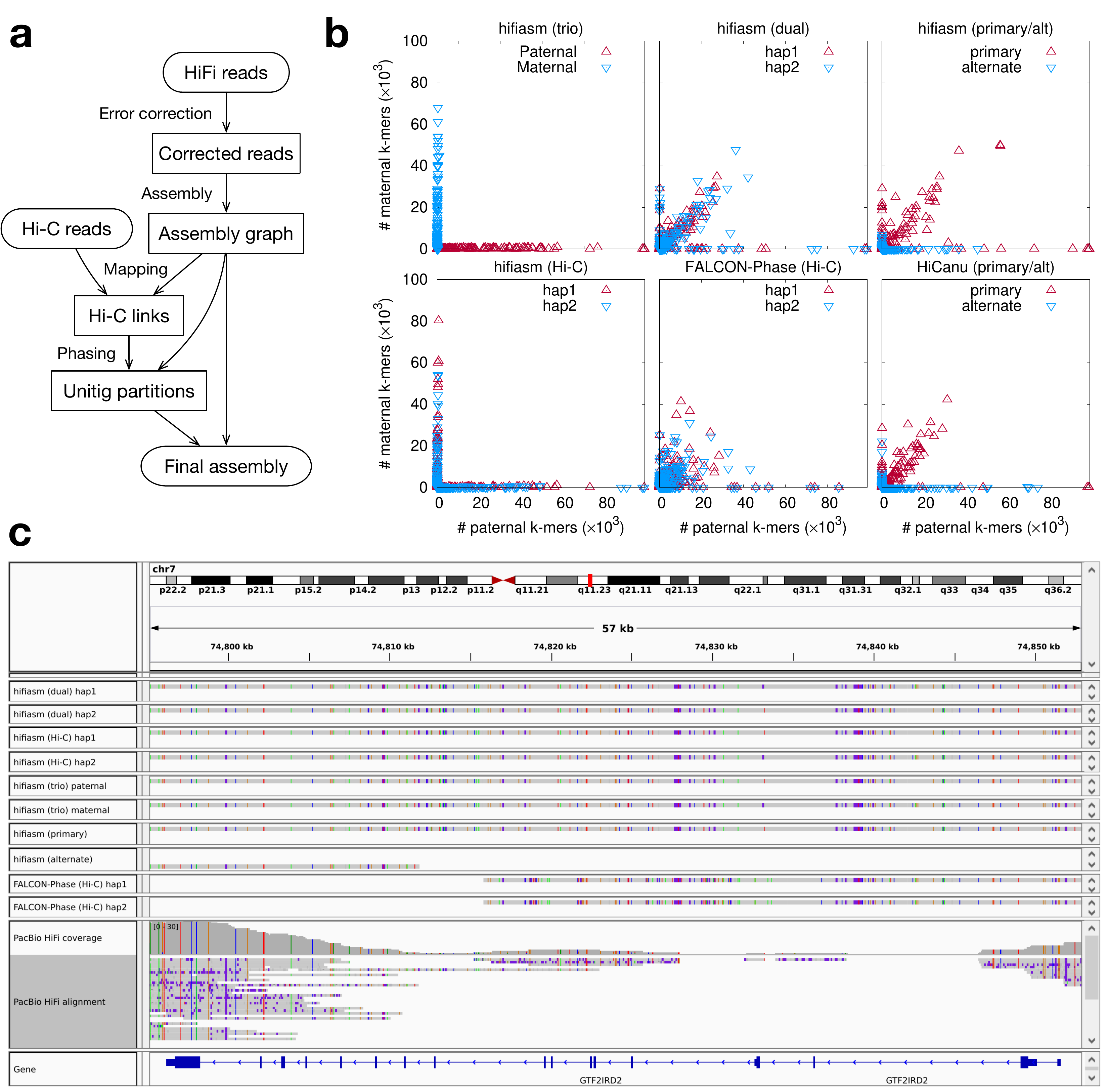}
\caption{{\bf Haplotype-resolved assembly using Hi-C data.} 
\textbf{(a)} Assembly workflow.
Hifiasm corrects reads and produces a phased assembly graph.
It then maps Hi-C short reads to the graph,
links unitigs in the assembly graph that share mapped Hi-C fragments,
and finds a bipartition of unitigs such that unitigs linked by many Hi-C fragments tend to be grouped together.
Hifiasm finally emits a haplotype-resolved assembly jointly considering the unitig partition and the assembly graph.
\textbf{(b)} Phasing accuracy of HG002 assemblies.
Each point corresponds to a contig.
Its coordinate gives the number of paternal- and maternal-specific 31-mers on the contig,
with these 31-mers derived from parental short reads.
Hifiasm (trio): haplotype-resolved hifiasm assembly with trio binning.
Hifiasm (dual): paired hifiasm assembly without Hi-C.
Hifiasm (primary/alt): primary and alternate hifiasm assembly without Hi-C.
Hifiasm (Hi-C): haplotype-resolved hifiasm assembly with Hi-C.
FALCON-Phase (Hi-C): FALCON-Phase assembly with Hi-C based on IPA contigs, 
acquired from its publication~\cite{kronenberg2021extended}.
HiCanu (primary/alt): primary and alternate HiCanu assembly without Hi-C.
All assemblies use the same HiFi and Hi-C datasets. 
\textbf{(c)} screenshot of contig and read alignment around gene \emph{GTF2IRD2}. 
}
\label{figp1}
\end{figure}

We also adapted our max-cut based phasing algorithm to HiFi-only data without Hi-C.
In this mode, hifiasm assumes within-haplotype sequence divergence being higher than between-haplotype divergence.
It effectively produces a pair of non-redundant primary assemblies representing a complete diploid genome.
We call such a pair of assemblies as a \emph{dual assembly}.
A dual assembly is often not haplotype-resolved because most long contigs are not haplotigs for a diploid sample of low heterozygosity.
Different from the traditional primary/alternate format,
a dual assembly is complete, more contiguous, more accurate and more useful in
downstream analysis as is shown below.
The idea of the dual assembly format was first introduced in a later version of Peregrine~\cite{chin2019human}, though our algorithm to derive the assembly is distinct.

\begin{table}[!tb]%\tiny
\captionsetup{singlelinecheck = false, justification=justified}
%\begin{adjustwidth}{-3.5in}{0in} % Comment out/remove adjustwidth environment if table fits in text column.
\footnotesize
\caption{Statistics of different assemblies}
%\scalebox{0.85}
{
\begin{tabular*}{\textwidth}{@{\extracolsep{\fill}} llcccccc}

\cline{1-8}

\multirow{2}{*}{Dataset} & \multirow{2}{*}{Assembler} & \makecell[c]{Size} & \makecell[c]{N50} & \makecell[c]{Hamming} & \makecell[c]{Multicopy genes} &  \multicolumn{2}{c}{\makecell[c]{Gene completeness}}\\
\cline{7-8}
                         &                            &           (Gb)               & (Mb)              & error (\%)              & missed (\%)              &  Complete (\%)          & Duplicated (\%) \\

\cline{1-8}
\multirow{3}{*}{\makecell[l]{HG002\\(HiFi + trio/Hi-C)
}}
& hifiasm (Hi-C)          & 2.927/3.037 & 57.6/44.9 & 0.89/0.99 & 18.23/26.51 & 99.08/99.24 & 0.32/0.33\\%& 97.99/98.20 & 0.37/0.34\\
& Falcon-Phase (Hi-C)     & 3.027/3.027 & 32.1/32.1 & 18.66/19.15 & 40.13/39.25 & 99.29/99.26 & 3.14/3.13 \\%98.12/98.07 & 3.00/3.00 \\
& hifiasm (trio)          & 2.913/3.024 & 57.2/53.1 & 0.80/0.75 & 24.60/19.03 & 98.99/99.22 & 0.28/0.34 \\%97.91/98.18 & 0.33/0.38\\
\cline{1-8}
\multirow{3}{*}{\makecell[l]{HG002\\(HiFi only)
}}
& hifiasm (dual)          & 3.018/3.047 & 46.3/56.2 & 24.00/25.39 & 18.95/20.78 & 99.06/99.05 & 0.36/0.36 \\%& 98.07/97.99 & 0.43/0.39\\
& hifiasm (primary/alt)   & 3.099/2.921 & 68.0/0.4 & 22.20/1.98 & 13.77/29.54 & 99.43/88.04 & 0.44/2.62 \\
& HiCanu (primary/alt)    & 2.960/3.143 & 48.4/0.3 & 27.76/0.68 & 34.95/20.62 & 98.88/85.63 & 0.19/5.15 \\%& 97.57/84.79 & 0.18/5.19\\
%& IPA (primary)          &  2.879 &  34.3 & 1.61 & 30.01 &  66.00 &  99.05 & 0.19 \\
\cline{1-8}
\multirow{4}{*}{\makecell[l]{HG00733 \\(HiFi + trio/Hi-C \\ /Strand-seq)
}}
& hifiasm (Hi-C)          & 3.016/3.056 & 41.7/42.7 & 1.34/1.43 & 16.00/19.51 & 99.49/99.46 & 0.33/0.27 \\%& 98.44/98.40 & 0.40/0.32\\
& DipAsm (Hi-C)           & 2.934/2.933 & 26.3/28.2 & 2.81/2.57 & 66.08/67.44 & 99.03/99.04 & 0.39/0.40 \\%& 97.14/97.00 & 0.31/0.32\\
& PGAS (Strand-seq)       & 2.905/2.900 & 30.1/25.9 & 3.25/2.60 & 66.48/68.31 & 99.15/99.18 & 0.16/0.15 \\%& 97.58/97.58 & 0.18/0.16\\
& hifiasm (trio)          & 3.039/3.017 & 41.7/44.1 & 0.80/1.00 & 15.29/19.67 & 99.52/99.28 & 0.30/0.30 \\%& 98.49/98.21 & 0.36/0.36\\
\cline{1-8}
\multirow{3}{*}{\makecell[l]{HG00733\\(HiFi only)
}}
& hifiasm (dual)          & 3.041/3.026 & 53.8/31.3 & 37.75/36.69 & 19.67/17.36 & 99.38/99.23 & 0.33/0.35 \\%& 98.28/98.21 & 0.46/0.46\\
& hifiasm (primary/alt)   & 3.081/3.013 & 59.1/0.3 & 40.14/2.23 & 12.34/28.66 & 99.53/85.06 & 0.46/2.91 \\%& 98.49/84.19 & 0.61/2.93\\
& HiCanu (primary/alt)    & 2.918/3.312 & 44.5/0.2 & 38.79/1.00 & 42.75/14.81 & 98.89/82.78 & 0.14/6.29 \\%& 97.48/81.98 & 0.14/6.45\\
%& IPA (primary)          &  2.861 &  34.6 &  0.72 & 41.85 &  66.28 &  99.16 &  0.16 \\
\cline{1-8}
%\multirow{2}{*}{\makecell{\textit{Meles meles} \\(fully phased)
\multirow{2}{*}{\makecell[l]{European badger \\(HiFi + trio/Hi-C)
}}
& hifiasm (Hi-C)          & 2.716/2.526 & 82.2/67.2 & 1.33/2.36 & & 96.74/94.49 & 1.78/1.64 \\
& hifiasm (trio)          & 2.619/2.587 & 84.4/65.8 & 0.68/3.26 & & 94.45/95.34 & 1.73/1.73 \\
\cline{1-8}
%\multirow{3}{*}{\makecell{\textit{Meles meles} \\(partial phased)
\multirow{3}{*}{\makecell[l]{European badger \\(HiFi only)
}}
& hifiasm (dual)          & 2.649/2.661 & 65.5/84.4 & 14.70/18.34 & & 95.81/96.56 & 1.85/1.78\\
& hifiasm (primary/alt)   & 2.738/1.692 & 84.5/0.2  & 12.43/1.89 & & 96.77/51.46 & 1.90/2.37\\
& HiCanu (primary/alt)    & 2.690/1.371 & 67.1/0.1 & 11.36/1.12 & & 96.75/38.30 & 1.96/2.57\\
%& IPA (primary)          &  2.587 &  47.8 & 1.29 & 12.33 &  & 96.74 &  2.33\\
\cline{1-8}
%\multirow{2}{*}{\makecell{\textit{Acipenser ruthenus} \\(fully phased)
\multirow{2}{*}{\makecell[l]{Sterlet \\(HiFi + trio/Hi-C)
}}
& hifiasm (Hi-C)          & 1.858/1.798 & 9.1/10.2 & 2.49/3.23 & & 93.35/93.27 & 60.16/60.33\\
& hifiasm (trio)          & 1.810/1.807 & 11.1/11.3 & 0.81/0.47 & & 93.76/93.74 &61.32/60.58  \\
\cline{1-8}
%\multirow{3}{*}{\makecell{\textit{Acipenser ruthenus} \\(partial phased)
\multirow{3}{*}{\makecell[l]{Sterlet \\(HiFi only)
}}
& hifiasm (dual)          & 1.847/1.836 & 8.9/10.2 & 11.36/11.87 & & 93.54/93.41 & 61.46/61.26 \\
& hifiasm (primary/alt)   & 1.922/1.884 & 27.4/1.5 & 25.29/0.92 & & 93.98/93.13 & 64.34/61.10\\
& HiCanu (primary/alt)    & 1.724/2.114 & 7.3/2.2 & 12.31/1.99 & & 91.48/90.25 & 42.47/59.97\\
%& IPA (primary)          &  1.263 & 7.33 &  0.99 & 20.85 &  & 89.18 & 16.15 \\
\cline{1-8}
%\multirow{2}{*}{\makecell{\textit{Porphyrio hochstetteri} \\(fully phased)
\multirow{2}{*}{\makecell[l]{South Island takahe \\(HiFi + trio/Hi-C)
}}
& hifiasm (Hi-C)          & 1.316/1.148 & 12.6/13.5 & 0.70/0.85 & & 97.03/90.49 & 0.52/0.48\\
& hifiasm (trio)          & 1.227/1.230 & 12.9/13.5 & 1.89/0.19 & & 91.40/92.07 & 0.50/0.54\\
\cline{1-8}
%\multirow{3}{*}{\makecell{\textit{Porphyrio hochstetteri} \\(partial phased)
\multirow{2}{*}{\makecell[l]{South Island takahe \\(HiFi only)
}}
& hifiasm (dual)          & 1.263/1.247 & 17.2/10.7 & 5.30/5.71 & & 95.25/93.10 & 0.49/0.49\\
& hifiasm (primary/alt)   & 1.319/0.643 & 16.3/0.3 & 5.46/0.92 & & 97.12/45.31 & 0.56/0.74\\
%& IPA (primary)          & 1.274 & 14.2 & 0.53 & 6.62 &  & 97.12 &  2.01\\
\cline{1-8}
%\multirow{2}{*}{\makecell{\textit{Diceros bicornis} \\(fully phased)
\multirow{2}{*}{\makecell[l]{Black Rhinoceros \\(HiFi + trio/Hi-C)
}}
& hifiasm (Hi-C)          & 2.927/3.053 & 29.5/28.8 & 1.33/1.37 & & 96.81/96.70 & 0.79/0.83\\
& hifiasm (trio)          & 2.979/3.005 & 30.2/28.8 & 0.95/0.34 & & 96.26/96.82 & 1.09/1.03\\
\cline{1-8}
%\multirow{3}{*}{\makecell{\textit{Diceros bicornis} \\(partial phased)
\multirow{3}{*}{\makecell[l]{Black Rhinoceros \\(HiFi only)
}}
& hifiasm (dual)          & 3.006/2.939 & 25.8/28.3 & 33.87/35.59 & & 95.59/95.17 & 0.78/0.87\\
& hifiasm (primary/alt)   & 3.089/2.791 & 41.4/0.8 & 37.23/3.31 & & 96.77/84.88 & 0.76/1.00\\
& HiCanu (primary/alt)    & 3.058/2.560 & 22.2/0.3 & 31.55/0.61 & & 96.79/70.11 & 1.53/1.38\\
%& IPA (primary)          & 2.831 & 21.7 & 1.08 & 32.78 &  & 96.94 & 6.13 \\
\cline{1-8}
\end{tabular*}
}

\begin{flushleft} \footnotesize{
All assemblies of the same sample use the same HiFi and Hi-C reads, except PGAS which relies on strand-seq data for phasing.
Each assembly consists of two sets of contigs.
The two sets may represent paternal/maternal with trio binning,
haplotype 1/haplotype 2 with haplotype-resolved assembly or hifiasm dual assembly,
or represent primary/alternate contigs.
The two numbers in each cell give the metrics for the two sets of contigs, respectively.
FALCON-Phase HG002 assembly, DipAsm and PGAS HG00733 assemblies were acquired from their associated publications. 
For South Island takahe, HiCanu could not produce assembly in 3 weeks so it is excluded.
The N50 of an assembly is defined as the sequence length of the shortest contig at 50\% of the total assembly size.
The completeness scores of all human assemblies were calculated by the asmgene method~\cite{li2016minimap} with GRCh38 as the reference genome.
The completeness of non-human assemblies were evaluated by BUSCO~\cite{simao2015busco}.
All samples have parental short reads.
The phasing switch error rates (Supplementary Table 1) and phasing hamming error rates were calculated with yak~\cite{cheng2021haplotype}.
The hamming error rate equals $\sum_i\min\{p_i,m_i\}/\sum_i(p_i+m_i)$ where $p_i$ and $m_i$ are the number of paternal- and maternal-specific 31-mers on contig $i$, respectively.
`Multicopy genes missed' is the percentage of multi-copy genes in GRCh38 (multiple mapping positions at $\ge$99\% sequence identity) that are not multi-copy in the assembly. 
This metric is only reported for human samples as other species lack high-quality reference genomes and good gene annotations.
}
\end{flushleft} \label{table_p1}

\end{table}

We first evaluated the phasing accuracy on the human HG002 dataset (Fig.~1b), taking trio phasing as the ground truth.
We confirm that hifiasm (Hi-C) or hifiasm (trio) contigs are largely haplotigs
with no contigs joining long stretches of paternal and maternal haplotypes.
Hifiasm (Hi-C) may put paternal and maternal contigs in one partition.
This is an innate ambiguity in Hi-C phasing as paternal and maternal chromosomes are indistinguishable in the cells of an offspring.
In the hifiasm and HiCanu primary/alternate assemblies (Fig.~1b), primary contigs are long but are not haplotigs; alternate contigs are haplotigs but are fragmented.
In comparison, both hap1 and hap2 assemblies in hifiasm (dual) behave like a primary assembly as is expected.

The advantage of hifiasm (Hi-C) and hifiasm (dual) is more apparent around segment duplications.
We examined the contig and read alignments around \emph{GTF2IRD2} (Fig.~1c), a key gene to the Williams-Beuren syndrome~\cite{makeyev2004gtf2ird2}.
This gene has a close paralog \emph{GTF2IRD2B} downstream.
Most HiFi reads from \emph{GTF2IRD2} are mapped to its paralog, leaving a coverage gap (Fig.~1c).
Alternate contigs have a similar issue.
Although FALCON-Phase Hi-C-based assemblies are not as fragmented as alternate contigs, it is still unable to resolve \emph{GTF2IRD2}.
Only haplotype-resolved and dual assemblies of hifiasm can go through this region on both haplotypes
and reveal the variations on this gene. 
Many challenging medically-relevant genes in segmental duplications like this gene can only be resolved by hifiasm~\cite{wagner2021towards}.

In comparison to other single-sample haplotype-resolved assembly pipelines
including FALCON-Phase~\cite{kronenberg2021extended}, DipAsm~\cite{garg2021chromosome} and PGAS~\cite{porubsky2021fully},
hifiasm Hi-C-based assemblies are more contiguous and have fewer phasing errors for human sample HG002 and HG00733 (Table~1).
%We can achieve comparable result at only 30-fold Hi-C coverage (Supplementary Table 2).
Hifiasm assemblies also miss fewer multi-copy genes, consistent with our earlier finding~\cite{cheng2021haplotype}.

We further applied hifiasm to several non-human datasets with parental data,
including two mammals (European badger, \textit{Meles meles} and Black rhinoceros, \textit{Diceros bicornis}
), a fish (Sterlet, \textit{Acipenser ruthenus}), and bird (South Island takahe, \textit{Porphyrio hochstetteri}).
The takahe is endangered, and the black rhinoceros is critically endangered, making it all the more imperative to obtain a more complete phased diploid assembly.

On most test datasets, single-sample Hi-C-based hifiasm is broadly comparable to trio-based hifiasm in terms of completeness, contiguity and phasing accuracy,
except for sterlet whose Hi-C-based assembly has a noticeably higher hamming error rate (Table~1).
We speculate this may be caused by residual tetraploidy whereby the ends of several chromosomes behave like a tetraploid organism~\cite{du2020sterlet}.
Such a genome organization would violate the diploid assumption of hifiasm Hi-C phasing and mislead the phasing to mix opposite parental sequences in one contig.
Hifiasm dual assemblies have similar contiguity and phasing switch error rates to the Hi-C assemblies on both haplotypes (Supplementary Table~1),
but their hamming error rates are higher as we are unable to correctly phase through regions of low heterozygosity with long reads alone.
Hi-C and dual assemblies aim to reconstruct both homologous haplotypes.
They will introduce a contig break if there is a random coverage drop on one of the two homologous haplotypes.
A primary assembly on the other hand only aims to reconstruct one haplotype.
If there is a coverage drop on one haplotype, it will switch to the other haplotype at higher coverage.
This makes primary assemblies less sensitive to uneven coverage and more contiguous.
Nonetheless, the focus of primary assembly on one haplotype greatly lowers the quality of the other haplotype.
The corresponding alternate contigs are often only $<$300kb in length and incomplete (Table 1).
A primary/alternate assembly does not represent a whole diploid genome.

The original hifiasm was the first assembler that could construct an assembly graph faithfully encoding the phasing of accurate long reads.
With the addition of Hi-C phasing on top of our earlier work,
hifiasm is so far the only assembler that can fully exploit the rich information in phased assembly graphs
and robustly produce unmatched high-quality haplotype-resolved assemblies for single individuals of various species in several hours (Supplementary Table 3).
Eliminating the barrier of the parental data requirement,
hifiasm has paved the way for population-scale haplotype-resolved assembly
which may shed light on the evolution and functionality of segmental duplications and complex gene families omitted in most analyses.

\section*{Methods}

~\\
\noindent \textbf{Overview of hifiasm assembly graphs.}
As is described in our earlier work~\cite{cheng2021haplotype}, a hifiasm assembly graph consists of unitigs with overlaps between them.
Unitigs have no phasing errors if generated correctly.
For a diploid sample, there are two types of unitigs: unitigs from heterozygous regions and unitigs from homozygous regions.
The two types can be distinguished based on their read coverage.
If we can determine the phase of each heterozygous unitig, we may use the existing hifiasm graph-binning algorithm
to derive a haplotype-resolved assembly by spelling contigs only composed of unitigs in the same phase.
The haplotype-resolved assembly is thus reduced to the unitig phasing problem (Supplementary Fig. 1).

~\\
\noindent \textbf{Producing a dual assembly with HiFi reads only.}
Recall that a dual assembly is a pair of non-redundant primary assemblies with each assembly representing a complete homologous haplotype.
At the overlapping step, hifiasm records \emph{trans} read overlaps, which are overlaps between reads from different homologous haplotypes.
Let $U_{st}$ be the number of \emph{trans} read overlaps between two unitigs $s$ and $t$.
It is a proxy to the similarity between the two unitigs.
For each heterozygous unitig $t$, let variable $\delta_t\in\{1,-1\}$ be its phase.
Hifiasm tries to maximize the following objective function to determine the phases of heterozygous unitigs:
\begin{equation}\label{eq:1}
F(\vec{\delta})=-\sum_{s,t} \delta_s\delta_t U_{st}
\end{equation}
where $\vec{\delta}$ is the vector of all unitig phases.
With this optimization, hifiasm tries to maximize the sequence similarity across the two phases, or equivalently, to minimize the similarity within each phase.
The objective function above takes a form similar to the Hamiltonian of Ising models and can be transformed to a graph maximum cut problem.
It can be approximately solved by a stochastic algorithm which will be explained later.
After determining the phases, hifiasm spells contigs composed of unitigs in the same phase.
This gives a dual assembly.
%With Hi-C data, hifiasm still performs a dual assembly internally to identify homologous relationship between unitigs.

~\\
\noindent \textbf{Mapping Hi-C reads to assembly graphs.}
Given a diploid sample, the total length of unitigs in the assembly graph approximately doubles the haploid genome size.
The majority of 31-mers in the graph have two copies and are not unique.
Unique 31-mers tend to harbor haplotype-specific heterozygous alleles.
Because for the purpose of phasing we only care about Hi-C read pairs that bridge two or more heterozygous alleles, we only index unique 31-mers in the assembly graph.
We map a Hi-C short read pair if it harbors two or more non-overlapping 31-mers and discard the remaining reads that are not informative to phasing.

Suppose a Hi-C short read is mapped to a heterozygous unitig $t$ with the algorithm above.
Such a mapping is a \emph{cis} mapping in that the Hi-C read is in the same phase as unitig $t$.
If HiFi reads around the Hi-C read mapping positions have \emph{trans} overlaps with HiFi reads on unitig $t'$,
we can infer the Hi-C read has a \emph{trans} mapping to $t'$, suggesting the Hi-C read is on the opposite phase of $t'$.
We need both \emph{cis} and \emph{trans} mappings for phasing.

~\\
\noindent \textbf{Modeling Hi-C phasing.}
For a Hi-C read pair $r$, let $x_{rt}=1$ if it has a \emph{cis} mapping to unitig $t$ or let $x_{rt}=-1$ if the Hi-C read has a \emph{trans} mapping to $t$;
otherwise $x_{rt}=0$.
Similar to the formulation of dual assembly, let $\delta_t\in\{1,-1\}$ denote the phase of a heterozygous unitig $t$.
$\{x_{rt}\}$ are observations while $\{\delta_t\}$ are variables whose values will be determined.

A Hi-C read pair may occasionally bridge two loci on different homologous haplotypes.
Such a Hi-C pair is called a \emph{trans} Hi-C pair.
Suppose a Hi-C read pair $r$ bridges unitig $s$ and $t$ and the probability of its being \emph{trans} is $\epsilon_r$.
By the definition of $x$, $x_{rs}$ and $x_{rt}$ can be either 1 or -1.
We have
$$
P(x_{rs},x_{rt}|\delta_s,\delta_t)=\left\{\begin{array}{ll}
(1-\epsilon_r)/2 & \mbox{if $x_{rt}x_{rs}\delta_t\delta_s=1$} \\
\epsilon_r/2 & \mbox{if $x_{rt}x_{rs}\delta_t\delta_s=-1$} \\
\end{array}\right.
$$
or equivalently
$$
P(x_{rs},x_{rt}|\delta_s,\delta_t)=\frac{1}{2}\sqrt{\epsilon_r(1-\epsilon_r)}\cdot\left(\frac{1-\epsilon_r}{\epsilon_r}\right)^{\frac{1}{2}x_{rt}x_{rs}\delta_t\delta_s}
$$
The composite log likelihood of $\vec{\delta}$ over all unitigs is
$$
\log\mathcal{L}(\vec{\delta})=\sum_r\sum_{s,t}\log P(x_{rs},x_{rt}|\delta_s,\delta_t)=C+\frac{1}{2}\sum_{s,t}\delta_s\delta_t\sum_r x_{rs}x_{rt}\log\frac{1-\epsilon_r}{\epsilon_r}
$$
where $C$ is not a function of $\delta$.
Define $w_r=\log \frac{1-\epsilon_r}{\epsilon_r}$, which is effectively the weight of Hi-C read pair $r$.
And introduce the \emph{cis} and \emph{trans} weights between unitigs:
$$
W_{st}=\sum_{r\in\{r|x_{rs}x_{rt}=1\}}w_r
$$
$$
\overline{W}_{st}=\sum_{r\in\{r|x_{rs}x_{rt}=-1\}}w_r
$$
The equation above can be written as
\begin{equation}\label{eq:2}
\log\mathcal{L}(\vec{\delta})=C+\frac{1}{2}\sum_{s,t}\delta_s\delta_t\left(W_{st}-\overline{W}_{st}\right)
\end{equation}
Maximizing the log-likelihood can be achieved by solving a max-cut problem again.
The derivation above extends reference-based phasing of the linker~\cite{Tourdot:2021aa} algorithm to unitig phasing.

As the probability of a Hi-C pair being \emph{trans} depends on the insert size~\cite{Edge:2017aa},
we calculate $\epsilon_r=\epsilon(d_r)$ where $d_r$ is the distance between the two mapping positions of Hi-C fragment $r$ on the unitig graph.
We use an empirical multiple rounds method to iteratively fit function $\epsilon(d)$, assuming all unitigs are haplotigs.
In each round of log-likelihood optimization, a \emph{trans} Hi-C read pair $r$ would be mapped to two heterozygous unitigs with different phases,
where their phases are determined by the last round of log-likelihood optimization.
We can then group Hi-C mappings around distance $d$ and calculate the fraction of \emph{trans} mappings in the group as $\hat{\epsilon}(d)$.
Note that in the initial round of optimization, all Hi-C read pairs are simply set to be \emph{cis} due to the lack of phase for each unitig.

~\\
\noindent \textbf{Finding near optimal solutions to the max-cut problem.}
Statistical physicists often optimize an equation like Eq.~(\ref{eq:1}) and~(\ref{eq:2}) with stochastic algorithms.
Hifiasm follows a similar route:
\begin{enumerate}
\item For each unitig $t$, randomly set $\delta_t$ to 1 or -1.
\item Arbitrarily choose a unitig $t$.
	Flip its phase (i.e. changing $\delta_t$ from 1 to -1 or vice versa) if doing so improves the objective function.
\item Repeat step 2 until the objective function cannot be improved.
	This reaches a local maximum.
	If the new local maximum is better than the best maximum so far, set it as the best maximum.
\item Perturb the best maximum either by randomly flipping a fraction of unitigs or by flipping all neighbors of a random unitig.
	Go to step 2 to look for a new local maximum.
\item Repeat steps 2 through 4 for 5,000 times by default and report the best local maximum in this process.
\end{enumerate}
In practical implementation at step 3, hifiasm may flip a pair of unitig sets at the same time
if the two sets of unitigs are inferred to come from opposite phases and to be homologous to each other.
Such pairs can be identified based on a ``bubble'' structure in the assembly graph and all-versus-all pairwise alignment between unitigs.
This heuristic speeds up convergence.
To reduce the effect of erroneous homologous unitig identification,
hifiasm does not apply the heuristic in the last round of optimization.

~\\
\noindent \textbf{Assembly of test samples.}
For all samples, we ran ``{\tt hifiasm hifi.fa.gz}'' to produce primary/alternate assemblies and dual assemblies,
and ran ``{\tt hifiasm --h1 hic\_1.fq.gz --h2 hic\_2.fq.gz hifi.fa.gz}'' to produce Hi-C based assemblies.
For trio binning assemblies, we used ``{\tt yak count -b37 -o parent.yak parent.fq.gz}'' to count k-mers for both parents
and ran ``{\tt hifiasm -1 father.yak -2 mother.yak hifi.fa}'' for assembly.
We ran HiCanu with ``{\tt canu genomeSize=\$GS useGrid=false -pacbio-hifi hifi.fa.gz}'',
where ``{\tt \$GS}'' is the genome size
which is set to 3.1g for human, 2.65g for European badger, 1.85g for Sterlet and 3.05g for Black rhinoceros.
We applied purge\_dups~\cite{guan2019identifying} to the initial HiCanu assemblies of HG002, HG00733 and sterlet as this improves their assemblies.
Purge\_dups and additional evaluation command lines can be found in the supplementary materials.

\section*{Acknowledgements}
This study was supported by US National Institutes of Health (grant R01HG010040,
U01HG010971 and U41HG010972 to H.L.). We thank members of the Vertebrate Genome 
Lab at the Rockefeller University and the Sanger genome team at the Sanger Institute for 
help with producing data for the non-human vertebrate species. Presentation and analyses of the 
completed reference genome assemblies will be reported on separately. We also thank the Human 
Pangenome Reference Consortium (HPRC) for making the HiFi and Hi-C data of HG002 and HG00733 publicly available. 
K-P.K. thanks the International Rhino Foundation for providing funding to generate the black rhinoceros assembly (grant number R-2018-1). 
The South Island takahe genome was funded by Revive and Restore and the University of Otago.
The South Island takahe reference genome was created in direct collaboration with the Takah\=e Recovery Team (Department of Conservation, New Zealand)
and Ng\=ai Tahu, the M\=aori kaitiaki ("guardians") of this taonga ("treasured") species.
Sequencing of the takahe genome was funded by Revive and Restore and the University of Otago.
L.U. was supported by a Feodor Lynen Fellowship of the Alexander von Humboldt Foundation, a Revive \& Restore Catalyst Science Fund, and the University of Otago.

\section*{Author contributions}
H.C. and H.L. designed the algorithm, implemented hifiasm and drafted the
manuscript. H.C. benchmarked hifiasm and other assemblers. E.D.J and O.F. 
coordinated generation of the non-human vertebrate species data, as part of the vertebrate genomes project.
K-P.K. sponsored the black rhinoceros genome. L.U. obtained the South Island takahe samples, all necessary permits, and funding for the South Island takahe reference genome.
L.U. and N.G. sponsored the South Island takahe genome.

\section*{Competing interests} 
H.L. is a consultant of Integrated DNA Technologies and on the Scientific Advisory Boards of
Sentieon and Innozeen.

\section*{Data availability} 
HG002 HiFi reads: SRR10382244, SRR10382245, SRR10382248 and SRR10382249;
Hi-C reads: ``{\tt HG002.HiC\_2*.fastq.gz}'' from \url{https://github.com/human-pangenomics/HG002\_Data\_Freeze\_v1.0};
Parental short reads: from the same HG002 data freeze.
HG00733 HiFi reads: ERX3831682;
Hi-C reads: SRR11347815;
parental short reads: ERR3241754 for HG00731 (father) and ERR3241755 for HG00732 (mother).
European badger (all data types): \url{https://vgp.github.io/genomeark/Meles\_meles/}.
Sterlet: \url{https://vgp.github.io/genomeark/Acipenser\_ruthenus/}.
South Island takahe: \url{https://vgp.github.io/genomeark/Porphyrio\_hochstetteri/}.
Black Rhinoceros: \url{https://vgp.github.io/genomeark/Diceros\_bicornis/}.
All evaluated assemblies are available at \url{ftp://ftp.dfci.harvard.edu/pub/hli/hifiasm-phase/v1/}.

\section*{Code availability} 
Hifiasm is available at
\url{https://github.com/chhylp123/hifiasm}.

\begin{comment}

\end{comment}

\bibliography{hifiasm}

\end{document}

\maketitle

\captionsetup{labelfont=bf}

\section{Software commands}

\subsection{Hifiasm}\label{sec:hifiasm}
To produce primary assemblies and alternate assemblies, hifiasm (version 0.15.5-r350) was
run with the following command:
\begin{quote}
\footnotesize\tt hifiasm -o \symbol{60}outputPrefix\symbol{62} -t \symbol{60}nThreads\symbol{62} --primary \symbol{60}HiFi-reads.fasta\symbol{62}
\end{quote}
For partial phased assemblies only with HiFi, hifiasm was run with the following command:
\begin{quote}
\footnotesize\tt hifiasm -o \symbol{60}outputPrefix\symbol{62} -t \symbol{60}nThreads\symbol{62} \symbol{60}HiFi-reads.fasta\symbol{62}
\end{quote}
For fully phased assemblies with Hi-C, hifiasm was run with the following command:
\begin{quote}
\footnotesize\tt hifiasm -o \symbol{60}outputPrefix\symbol{62} --h1 \symbol{60}HiC-reads-R1.fasta\symbol{62} --h2 \symbol{60}HiC-reads-R2.fasta\symbol{62} -t \symbol{60}nThreads\symbol{62} \symbol{60}HiFi-reads.fasta\symbol{62}
\end{quote}
For trio-binning assembly, we first built the paternal trio index and the
maternal trio index by yak (version 0.1-r62-dirty) with the following commands:   
\begin{quote}
\footnotesize\tt yak count -b37 -t \symbol{60}nThreads\symbol{62} -o \symbol{60}pat.yak\symbol{62} \symbol{60}paternal-short-reads.fastq\symbol{62}\\
\footnotesize\tt yak count -b37 -t \symbol{60}nThreads\symbol{62} -o \symbol{60}mat.yak\symbol{62} \symbol{60}maternal-short-reads.fastq\symbol{62}
\end{quote}
and then we produced the paternal assembly and the maternal assembly with the
following command:   
\begin{quote}
\footnotesize\tt hifiasm -o \symbol{60}outputPrefix\symbol{62} -t \symbol{60}nThreads\symbol{62} -1 \symbol{60}pat.yak\symbol{62} -2 \symbol{60}mat.yak\symbol{62} \symbol{60}HiFi-reads.fasta\symbol{62}
\end{quote}

\subsection{HiCanu}
For primary assembly, HiCanu (version 2.1.1) was run with the following command
line:
\begin{quotation}
\footnotesize\tt\noindent
canu -p asm -d \symbol{60}outDir\symbol{62} genomeSize=\symbol{60}GSize\symbol{62} useGrid=false maxThreads=\symbol{60}nThreads\symbol{62} \symbol{92} \\
\indent -pacbio-hifi \symbol{60}HiFi-reads.fasta\symbol{62}
\end{quotation}
The contigs labeled by `\texttt{suggestedBubbles=yes}' were removed from the
primary assembly. We ran purge\_dups (version v1.2.5) to postprocess the HG002, HG00733 
and Sterlet assemblies. The HiCanu assemblies of European badger and Black 
Rhinoceros were not purged as purge\_dups resulted in significantly worse assemblies.

\subsection{Purge\_dups}
Purge\_dups (version 1.2.5) was used to postprocess the output primary
assemblies of HiCanu for HG002, HG00733 and Sterlet. The commands are as follows:   
\begin{quote}
\footnotesize\tt
minimap2 -I10G -xmap-pb \symbol{60}asm.fa\symbol{62} \symbol{60}HiFi-reads.fasta\symbol{62} -t \symbol{60}nThreads\symbol{62} \symbol{62} \symbol{60}read-aln.paf\symbol{62} \\
bin/pbcstat \symbol{60}read-aln.paf\symbol{62} \\
bin/calcuts PB.stat \symbol{62} cutoffs \\
bin/split\_fa \symbol{60}asm.fa\symbol{62} \symbol{62} \symbol{60}split.fa\symbol{62} \\
minimap2 -I10G -xasm5 -DP \symbol{60}split.fa\symbol{62} \symbol{60}split.fa\symbol{62} -t \symbol{60}nThreads\symbol{62} \symbol{62} \symbol{60}ctg-aln.paf\symbol{62} \\
bin/purge\_dups -2 -T cutoffs -c PB.base.cov \symbol{60}ctg-aln.paf\symbol{62} \symbol{62} \symbol{60}dups.bed\symbol{62} \\
bin/get\_seqs \symbol{60}dups.bed\symbol{62} \symbol{60}asm.fa\symbol{62}
\end{quote}
We then manually adjusted the cutoff thresholds of purge\_dups as ``{\tt 5 7 11 30 22 42}'', 
``{\tt 5 7 11 30 22 42}'' and ``{\tt 5 24 24 25 25 92}'' for HG002, HG00733 and Sterlet, respectively.

\subsection{Running asmgene}
For HG00733 and HG002, We aligned the cDNAs to the GRCh38 human reference genome and contigs by minimap2 (version 2.20-r1061), and evaluated the gene completeness with paftools.js from the minimap2 package:
\begin{quote}
\tt\footnotesize minimap2 -cxsplice:hq -t \symbol{60}nThreads\symbol{62} \symbol{60}asm.fa\symbol{62} \symbol{60}cDNAs.fa\symbol{62} \symbol{62} \symbol{60}aln.paf\symbol{62} \\
\tt\footnotesize paftools.js asmgene -a -i.97 \symbol{60}ref.paf\symbol{62} \symbol{60}asm.paf\symbol{62}
\end{quote}
When evaluating multi-copy genes missed in an assembly, we replaced `{\tt
-i.97}' to `{\tt -i.99}' to increase the resolution.

\subsection{BUSCO}
BUSCO (version 5.1.3) was used with the following command:   
\begin{quote}
\tt\footnotesize busco -i \symbol{60}asm.fa\symbol{62} -m genome -o \symbol{60}outDir\symbol{62} -c \symbol{60}nThreads\symbol{62} -l \symbol{60}lineage\_dataset\symbol{62}
\end{quote}
where `\texttt{lineage\_dataset}' was set to \emph{mammalia\_odb10} for European badger and Black Rhinoceros, set to \emph{actinopterygii\_odb10} for Sterlet and set to \emph{aves\_odb10} for South Island takahe.

\subsection{QV evaluation}
We used yak (version 0.1-r62-dirty) to measure the per-base consensus accuracy (QV). To this end, we built
the index for the short reads coming from the same sample and did evaluation: 
\begin{quote}
\tt\footnotesize
yak count -b37 -t \symbol{60}nThreads\symbol{62} -o \symbol{60}sr.yak\symbol{62} \symbol{60}short-reads.fastq\symbol{62} \\
yak qv -t \symbol{60}nThreads\symbol{62} \symbol{60}sr.yak\symbol{62} \symbol{60}asm.fa\symbol{62}
\end{quote}

\subsection{IGV visualization}
We used IGV (version 2.10.0) to visualize alignment around complex genes on the GRCh38 human reference genome.
HiFi reads were aligned by minimap2 (version 2.20-r1061) as follows: 
\begin{quote}
\tt\footnotesize minimap2 -ax map-hifi -t \symbol{60}nThreads\symbol{62} \symbol{60}ref.fa\symbol{62} \symbol{60}HiFi-reads.fasta\symbol{62}
\end{quote}
Assemblies were aligned with the following command:   
\begin{quote}
\tt\footnotesize minimap2 -ax asm5 -t \symbol{60}nThreads\symbol{62} \symbol{60}ref.fa\symbol{62} \symbol{60}asm.fa\symbol{62}
\end{quote}
%When evaluating LPA gene, we replaced `{\tt asm5}' to `{\tt asm20}' as LPA of HG002 is more divergent in comparison to GRCh38.

\begin{comment}
\newpage

\section{Estimating Quality Values}

Let $x_i$ be the number of $k$-mers occurring $i$ times in short reads. For a
$k$-mer occurring $i$ times in short reads, we add its count in contigs to
$y_i$. In particular, $y_0$ is the number of $k$-mers in contigs that do not
appear in short reads. Then $N\triangleq\sum_{i=0} y_i$ gives the total number
of $k$-mers in contigs. If we assume the set of true $k$-mers is the same as
the set of $k$-mers inferred from short reads, the maximum-likelihood estimate
of the per-base error rate of contigs is
$$
\hat{\epsilon}=1-\left(1-\frac{y_0}{N}\right)^{1/k}
$$
By definition, the quality value is $-10\log_{10}\hat{\epsilon}$.

In practice, however, there are sequencing errors in short reads, which leads
to underestimated $y_0$ and thus overestimated QV. Meanwhile, $k$-mers from
short reads may be incomplete, which leads to underestimated QV. Yak attempts
to correct both effects.

Suppose each false $k$-mer in short reads may occur in contigs at probability
$p$. Let 
$$
a=\argmin_{i>0} y_i
$$
$$
b=\argmax_{i>0} y_i
$$
$$
c=y_b / x_b
$$
Compute
$$
e_i=\frac{x_i-y_i/c}{1-p}
$$
$$
y'_i=y_i-e_i\cdot c\cdot p
$$
\end{comment}

\newpage

\begin{table}[!tb]%\tiny
\captionsetup{singlelinecheck = false, justification=justified}
%\begin{adjustwidth}{-3.5in}{0in} % Comment out/remove adjustwidth environment if table fits in text column.
\footnotesize
\caption{Statistics of different assemblies}
%\scalebox{0.85}
{
\begin{tabular*}{\textwidth}{@{\extracolsep{\fill}} llcc}

\cline{1-4}

\multirow{2}{*}{Dataset} & \multirow{2}{*}{Assembler} & \multirow{2}{*}{QV} & \makecell[c]{Switch}\\

                         &                            &                          & error (\%) \\

\cline{1-4}
\multirow{3}{*}{\makecell[l]{HG002\\(HiFi + trio/Hi-C)
}}
& hifiasm (Hi-C)        & 52.28/51.86 & 0.91/0.90\\

& FALCON-Phase (Hi-C)        &  43.43/43.55 & 1.52/1.48\\

& hifiasm (trio)        & 52.28/51.91 & 0.88/0.99\\

\cline{1-4}
\multirow{3}{*}{\makecell[l]{HG002\\(HiFi only)
}}
& hifiasm (dual)        & 51.73/52.04 & 1.00/1.06\\

& hifiasm (primary/alt)        & 51.86/ 51.47 & 0.90/0.91\\

& HiCanu (primary/alt)        & 52.15/38.08 & 1.20/0.70\\

\cline{1-4}

\multirow{4}{*}{\makecell[l]{HG00733 \\(HiFi + trio/Hi-C \\ /Strand-seq)
}}
& hifiasm (Hi-C)        & 49.97/50.45 & 1.06/1.09\\

& DipAsm (Hi-C)        & 41.57/41.65 & 2.30/2.15\\

& PGAS (Strand-seq)        & 45.52/46.06 & 1.44/1.63\\

& hifiasm (trio)        & 50.19/50.60 & 0.98/0.98\\
\cline{1-4}
\multirow{3}{*}{\makecell[l]{HG00733\\(HiFi only)
}}
& hifiasm (dual)        & 50.09/50.60 & 1.34/1.35\\

& hifiasm (primary/alt)        & 50.24/49.10 & 1.30/0.91\\

& HiCanu (primary/alt)        & 50.84/36.60 & 1.38/0.99\\

\cline{1-4}

\multirow{2}{*}{\makecell[l]{European badger \\(HiFi + trio/Hi-C)
}}
& hifiasm (Hi-C)          & & 1.05/2.01 \\
& hifiasm (trio)          & & 0.85/2.71 \\
\cline{1-4}
\multirow{3}{*}{\makecell[l]{European badger \\(HiFi only)
}}
& hifiasm (dual)          & & 1.29/1.62\\
& hifiasm (primary/alt)   & & 1.11/1.41\\
& HiCanu (primary/alt)    & & 1.11/1.34\\
\cline{1-4}
\multirow{2}{*}{\makecell[l]{Sterlet \\(HiFi + trio/Hi-C)
}}
& hifiasm (Hi-C)          & & 0.85/0.84\\
& hifiasm (trio)          & & 1.09/0.60\\
\cline{1-4}
\multirow{3}{*}{\makecell[l]{Sterlet \\(HiFi only)
}}
& hifiasm (dual)          & & 0.85/0.84\\
& hifiasm (primary/alt)   & & 0.83/0.88\\
& HiCanu (primary/alt)    & & 0.82/0.93\\
\cline{1-4}
\multirow{2}{*}{\makecell[l]{South Island takahe \\(HiFi + trio/Hi-C)
}}
& hifiasm (Hi-C)          & & 0.37/0.53\\
& hifiasm (trio)          & & 1.67/0.19\\
\cline{1-4}
\multirow{2}{*}{\makecell[l]{South Island takahe \\(HiFi only)
}}
& hifiasm (dual)          & & 0.41/0.46\\
& hifiasm (primary/alt)   & & 0.42/0.34\\
\cline{1-4}
\multirow{2}{*}{\makecell[l]{Black Rhinoceros \\(HiFi + trio/Hi-C)
}}
& hifiasm (Hi-C)          & & 0.72/0.64\\
& hifiasm (trio)          & & 0.83/0.46\\
\cline{1-4}
\multirow{3}{*}{\makecell[l]{Black Rhinoceros \\(HiFi only)
}}
& hifiasm (dual)          & & 0.76/0.77\\
& hifiasm (primary/alt)   & & 0.75/0.62\\
& HiCanu (primary/alt)    & & 0.87/0.58\\
\cline{1-4}
\end{tabular*}
}

\begin{flushleft} \footnotesize{
Each assembly consists of two sets of contigs.
The two sets represent paternal/maternal with trio binning,
haplotype 1/haplotype 2 with haplotype-resolved assembly or hifiasm dual assembly,
or represent primary/alternate contigs.
The two numbers in each cell give the metrics for the two sets of contigs, respectively.
QV is the Phred-scaled contig base error rate measured by comparing 31-mers in contigs to 31-mers in short reads from the same sample. QV was not reported for non-human samples as there are no short reads from the same sample. 
The phasing switch error rate is the percentage of adjacent parental-specific 31-mer pairs that come from different parental haplotypes. It was calculated with yak.}
\end{flushleft} \label{table_sp1}

\end{table}

\begin{table}[!h]%\tiny
\captionsetup{singlelinecheck = false, justification=justified}
%\begin{adjustwidth}{-3.5in}{0in} % Comment out/remove adjustwidth environment if table fits in text column.
\footnotesize
\caption{Statistics of HG002 hifiasm (Hi-C) assemblies with different coverages of Hi-C}
%\scalebox{0.85}
{
\begin{tabular*}{\textwidth}{@{\extracolsep{\fill}} llccccccc}

\cline{1-9}

\multirow{2}{*}{Dataset} & \multirow{2}{*}{Haplotype} & \makecell[c]{Size} & \makecell[c]{N50} & \makecell[c]{Switch} & \makecell[c]{Hamming} &  \makecell[c]{Multicopy genes} & \multicolumn{2}{c}{\makecell[c]{Completeness (asmgene or BUSCO)}}\\
\cline{8-9}
                         &                            &           (Gb)               & (Mb)              & error (\%)              & error (\%)     & missed  (\%)         &  Complete (\%)          & Duplicated (\%) \\

\cline{1-9}
\multirow{2}{*}{\makecell{30x Hi-C}}
& haplotype 1        &  2.927 & 57.6 & 0.91 & 0.89 & 18.23 & 99.08 & 0.32\\

& haplotype 2        & 3.037 & 44.9 & 0.90 & 0.99  & 26.51 & 99.24 & 0.33\\
\cline{1-9}
\multirow{2}{*}{\makecell{60x Hi-C}}
& haplotype 1        & 3.039 & 55.1 & 0.89 & 1.05 & 18.55 & 99.08 & 0.32\\

& haplotype 2        & 2.928 & 59.1 & 0.88 & 1.03 & 23.81 & 99.28 & 0.30\\
\cline{1-9}
\multirow{2}{*}{\makecell{90x Hi-C}}
& haplotype 1        &  2.941 & 68.0 & 0.86 & 0.87  & 23.49 & 99.07 & 0.31\\

& haplotype 2        &  3.027 & 46.9 & 0.89 & 0.96 & 21.42 & 99.27 & 0.32\\

\cline{1-9}

\end{tabular*}
}

\label{table_sp2}

\end{table}

%\makeatletter
%\addtocounter{table}{32}
%\makeatother
\begin{table}[!h]
\captionsetup{singlelinecheck = false, justification=justified}
%\begin{adjustwidth}{-3.5in}{0in} % Comment out/remove adjustwidth environment if table fits in text column.
\footnotesize
\caption{Run time and peak memory usage of different assemblers}
{
\begin{tabular*}{\textwidth}{@{\extracolsep{\fill}} llccccc}

\hline
\multirow{2}{*}{Dataset}&\multirow{2}{*}{Metric}&hifiasm&hifiasm&hifiasm&hifiasm&HiCanu\\
&&(Hi-C)&(trio)&(dual)& (primary/alt) & (primary/alt)\\

\hline
\multirow{3}{*}{HG002}
& Elapsed time (h) & 10.8 & 10.1 & 9.5 &9.4 &48.5\\
& CPU time (h) & 366.9 & 368.3 & 358.1 & 357.2 &1,369.3\\
& Peak Memory (Gb) & 149.5 & 142.9 & 142.9 & 142.9 & 83.7\\

\hline
\multirow{3}{*}{HG00733}
& Elapsed time (h) & 9.1 & 8.6 & 7.7 &  7.7 & 43.9\\
& CPU time (h) & 292.2 & 293.7  & 282.0 &  281.9 & 1,202.2\\
& Peak Memory (Gb) & 147.1 & 134.9 & 134.9 & 134.9 &70.8\\

\hline
\multirow{3}{*}{European badger}
& Elapsed time (h) & 13.2 & 11.7 & 10.9 & 10.9 & 65.3\\
& CPU time (h) & 466.3 & 452.5 & 440.2 & 440.1 & 2,185.1\\
& Peak Memory (Gb) & 178.8 & 178.8 & 178.8 & 178.8 & 131.2\\

\hline
\multirow{3}{*}{Sterlet}
& Elapsed time (h) & 14.5 & 13.6 & 13.1 & 13.0 & 186.6\\
& CPU time (h) & 565.4 & 570.1 & 551.8 & 551.7 & 7,701.7\\
& Peak Memory (Gb) & 184.1 & 184.1 & 184.1 & 184.1 &41.2\\

\hline
\multirow{3}{*}{South Island takahe}
& Elapsed time (h) & 4.3 & 3.9 & 3.7 & 3.6 & \\
& CPU time (h) & 149.1 & 147.3 & 143.2 & 143.0 & \\
& Peak Memory (Gb) & 70.0 & 66.1 & 66.1 & 66.1 & \\

\hline
\multirow{3}{*}{Black Rhinoceros}
& Elapsed time (h) & 13.7 & 12.2 & 11.2 & 11.2 & 41.9\\
& CPU time (h) & 461.5 & 451.7 & 436.0 & 435.7 & 1,358.5\\
& Peak Memory (Gb) & 194.5 & 194.5 & 194.5 & 194.5 & 42.6\\

\hline
\end{tabular*}
}

\begin{flushleft} \footnotesize{
All assemblies were generated using the same machine with 48 CPU threads. 
For HiCanu (primary/alt), we ran purge\_dups to postprocess the HG002, HG00733 
and Sterlet assemblies. 
The HiCanu assemblies of European badger and Black Rhinoceros were not purged as 
purge\_dups resulted in significantly worse assemblies. 
For South Island takahe, HiCanu could not produce assembly in 3 weeks.}
\end{flushleft} \label{table_sp3}

\end{table}

\newpage
\clearpage

\begin{figure}[!h]
\centering
\includegraphics[width=1\textwidth]{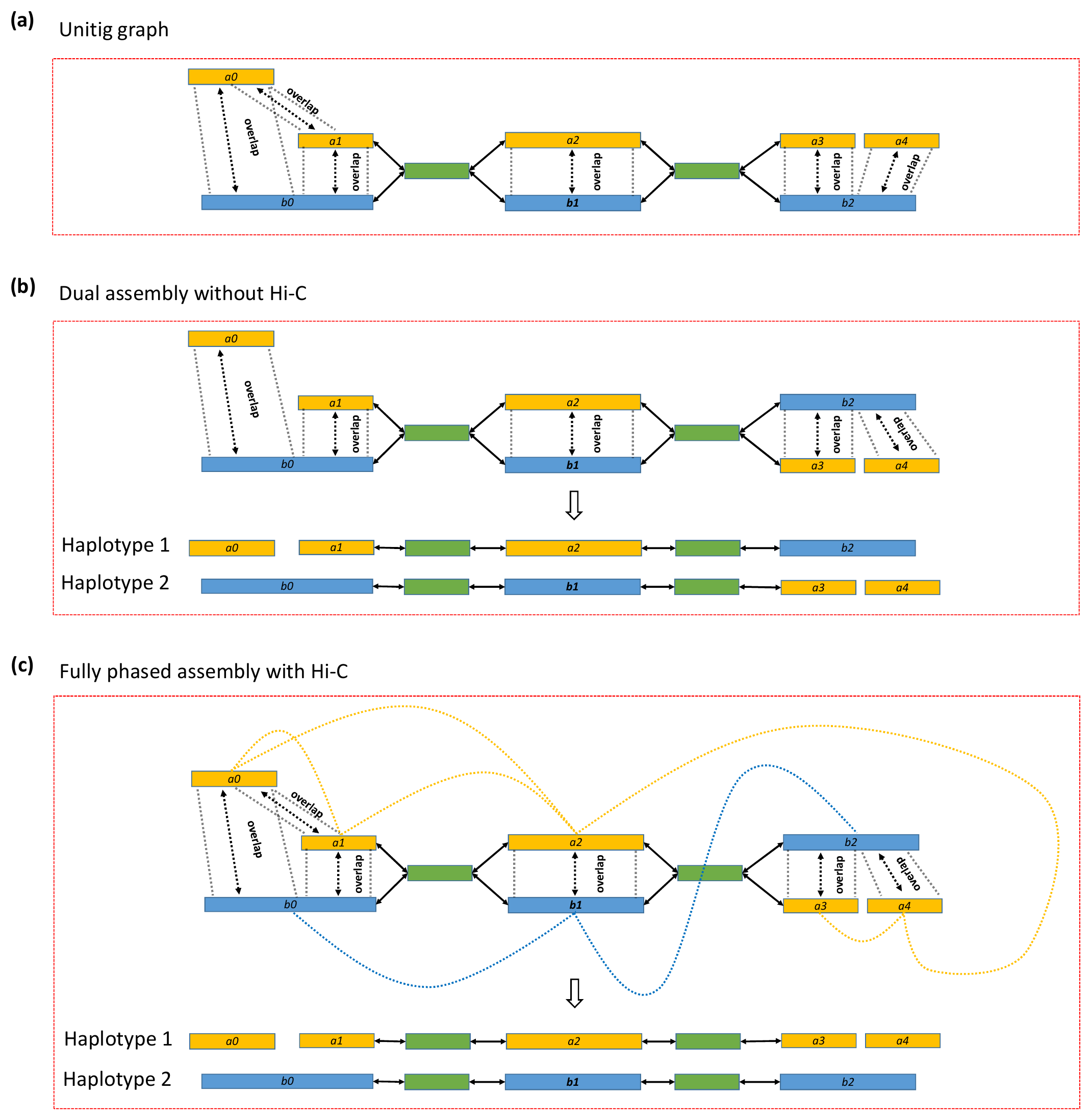}
\caption{{\bf Outline of the hifiasm phasing algorithm.} \textbf{(a)} 
Rectangles in orange and blue represent heterozygous unitigs (haplotigs) coming from
haplotype 1 and haplotype 2, respectively. Rectangles in green are homozygous unitigs. 
Solid lines between unitig ends indicate that the corresponding unitigs are linked in unitig graph. 
By utilizing all-to-all overlap calculation, hifiasm identifies 6 overlaps between unitigs: 
$a0 \leftrightarrow a1$, $a0 \leftrightarrow b0$, $a1 \leftrightarrow b0$, $a2 \leftrightarrow b1$, 
$a3 \leftrightarrow b2$ and $a4 \leftrightarrow b2$.
\textbf{(b)} Without Hi-C, the optimization model for dual assembly identify 
$a0 \leftrightarrow a1$ does not represent corresponding homologous regions at different haplotypes. 
Hifiasm phasing algorithm bins homologous unitigs to different haplotypes to produce two non-redundant primary assemblies.
\textbf{(c)} Dashed lines represent Hi-C contacts between unitigs. Two fully phased assemblies are generated by 
integrating Hi-C data.}
\label{figf2}
\end{figure}

\begin{comment}
\begin{figure}[!t]
\subfloat[]{\includegraphics[width=1\textwidth]{LPA_alignment_overview.png}}\\
\subfloat[]{\includegraphics[width=1\textwidth]{LPA_alignment_detail.png}}\\
\caption{{\bf IGV screenshot around segmentally duplicated gene $LPA$ of HG002.} 
Hifiasm (dual), hifiasm (Hi-C) and hifiasm (trio) assemblies completely resolve this challenging 
gene. In contrast, it is hard to call $LPA$ with mapping-based methods due to the read mapping
issues.
\textbf{(a)} Overview of assembly alignments and HiFi read alignments around $LPA$.
\textbf{(b)} Alignment details around $LPA$.
}
\label{figf3}
\end{figure}
\end{comment}
\bibliography{hifiasm}      % Bibliography file (usually '*.bib' )